\documentclass[twocolumn]{aastex62}
\usepackage{amsmath}

\newcommand{\srcfull}{FIRST~J141918.9+394036}
\newcommand{\src}{J1419+3940}

\shorttitle{Discovery of Radio Transient FIRST~J141918.9+394036}
\shortauthors{Law et al.}

\begin{document}

\title{Discovery of the Luminous, Decades-Long, Extragalactic Radio Transient FIRST~J141918.9+394036}

\author[0000-0002-4119-9963]{C.J.~Law}
\affiliation{Department of Astronomy and Radio Astronomy Lab, University of California, Berkeley, CA 94720, USA}
\affiliation{Dunlap Institute for Astronomy and Astrophysics, University of Toronto, 50 St. George Street, Toronto, Ontario, M5S 3H4, Canada}
\correspondingauthor{Casey J.~Law}
\email{claw@astro.berkeley.edu}


\author[0000-0002-3382-9558]{B.M.~Gaensler}
\affiliation{Dunlap Institute for Astronomy and Astrophysics, University of Toronto, 50 St. George Street, Toronto, Ontario, M5S 3H4, Canada}
\affiliation{Department of Astronomy and Astrophysics, University of Toronto, 50 St. George Street, Toronto, Ontario, M5S 3H4, Canada}

\author[0000-0002-4670-7509]{B.D.~Metzger}
\affiliation{Department of Physics and Columbia Astrophysics Laboratory, Columbia University, New York, NY 10027, USA}

\author{E.O.~Ofek}
\affiliation{Benoziyo Center for Astrophysics, Weizmann Institute of Science, 76100 Rehovot, Israel}

\author[0000-0002-5951-0756]{L.~Sironi}
\affiliation{Department of Astronomy, Columbia University, New York, NY 10027, USA}

\begin{abstract}
We present the discovery of a slowly-evolving, extragalactic radio transient, FIRST J141918.9+394036, identified by comparing a catalog of radio sources in nearby galaxies against new observations from the Very Large Array Sky Survey. Analysis of other archival data shows that FIRST J141918.9+394036 faded by a factor of $\sim$50 over 23 years, from a flux of $\sim$26 mJy at 1.4 GHz in 1993 to an upper limit of 0.4 mJy at 3 GHz in 2017. FIRST J141918.9+394036 is likely associated with the small star-forming galaxy SDSS J141918.81+394035.8 at a redshift $z=0.01957$\ ($d=87$ Mpc), which implies a peak luminosity $\nu L_\nu \gtrsim 3\times10^{38}$ erg s$^{-1}$. If interpreted as an isotropic synchrotron blast wave, the source requires an explosion of kinetic energy $\sim10^{51}$ erg some time prior to our first detection in late 1993. This explosion could plausibly be associated with a long gamma-ray burst (GRB) or the merger of two neutron stars. Alternatively, FIRST J141918.9+394036 could be the nebula of a newly-born magnetar. The radio discovery of any of these phenomena would be unprecedented. Joint consideration of the event light curve, host galaxy, lack of a counterpart gamma-ray burst, and volumetric rate suggests that FIRST J141918.9+394036 is the afterglow of an off-axis (``orphan'') long GRB. The long time baseline of this event offers the best available constraint in afterglow evolution as the bulk of shock-accelerated electrons become non-relativistic. The proximity, age, and precise localization of FIRST J141918.9+394036 make it a key object for understanding the aftermath of rare classes of stellar explosion.
\end{abstract}


\section{Introduction}
\label{sec:intro}

Multiple large radio interferometric surveys were conducted in the 1990s with sufficiently high spatial resolution to facilitate comparison to optical surveys.  These first robust statistical samples enabled novel tests of the physics of high-redshift quasars \citep{2000ApJ...538...72B}, star-forming galaxies \citep{2004ApJS..154..147A}, Galactic pulsars \citep{2000ApJ...529..859K}, constraints on the beaming factors of gamma-ray bursts (GRBs) \citep{2006ApJ...639..331G}, and more.  Today, a new generation of high-energy, optical, and radio sky surveys are being conducted with a focus on sensitivity to time-domain phenomena like supernovae and GRBs \citep{2004ApJ...611.1005G, 2009PASP..121.1395L, 2016ApJ...818..105M}. The Karl G.~Jansky Very Large Array (VLA) is in the midst of a sky survey whose design explicitly supports transient science \citep[``VLASS'';][]{vlass}.

Transient surveys at optical and high-energies have revolutionized our understanding of relativistic transients like GRBs \citep{2006ARA&A..44..507W} and tidal disruption events \citep[TDEs;][]{2011Sci...333..203B,2012Natur.485..217G}, but they are less sensitive to the vast majority of events not beamed in our direction \citep{2001ApJ...562L..55F}. In contrast, radio emission traces the total kinetic energy ({\it calorimetry}) of the interaction of ejecta with the interstellar medium.  This makes radio observations valuable for transient discovery and unbiased rate estimates \citep{2005ApJ...619..994F}. Even non-relativistic explosions, such as the hypothesized magnetar-powered supernovae \citep{2010ApJ...717..245K}, may produce radio emission at late times, once the supernova ejecta become transparent to free-free absorption and the birth nebula of the inner engine is revealed \citep{2016MNRAS.461.1498M,2017ApJ...841...14M,2017ApJ...850...55N,2018MNRAS.474..573O}. The recent association of a fast radio burst \citep[FRB;][]{2017Natur.541...58C} with a luminous, persistent radio source has suggested that such slowly-evolving radio transients may provide signposts for the discovery of past energetic events giving birth to FRB-sources \citep{2017ApJ...839L...3K, 2018arXiv180605690M}.

Tremendous effort has been invested in blind radio transient surveys \citep{2010ApJ...719...45C, 2010ApJ...725.1792B, 2011MNRAS.415....2B, 2011ApJ...740...65O, 2013ApJ...768..165M,  2016MNRAS.458.3506R, 2017MNRAS.466.1944M}, but the requirements are severe. For example, orphan-GRB afterglows require a search at mJy-sensitivity, over $10^4$~deg$^2$, with multiple epochs over many years \citep{2002ApJ...576..923L, 2015ApJ...806..224M}. This class of radio survey is only becoming available today by comparison of VLA surveys from the 1990s \citep{1995ApJ...450..559B, 1998AJ....115.1693C} to the VLASS.

\citet{2017ApJ...846...44O} collected a sample of galaxies with luminosity distance smaller than 108\,Mpc ($z=0.025$) and compared them to point sources in the VLA FIRST survey \citep{1995ApJ...450..559B}. He identified a set of radio sources with potential association to nearby galaxies and relatively high luminosities ($\nu L_\nu > 3\times 10^{37}$~erg~s$^{-1}$). Here, we describe the analysis of new data from VLASS and archival radio data that reveals that one of these sources, \srcfull\ (hereafter \src), is a luminous radio transient. \S\ref{sec:dis} describes how we discovered the transient in VLASS and \S\ref{sec:cha} presents a radio light curve compiled from multiple telescopes with detections spanning more than two decades. We also describe the properties of the host galaxy of \src, constraints on gamma-ray emission, and an estimate of the volumetric rate for \src-like transients. \S\ref{sec:ori} presents calculations and modeling of the radio data that suggests that \src\ is either the afterglow of an orphan GRB, or is the wind nebula produced in the aftermath of a magnetar powered supernova. We summarize the results and present future tests of this model in \S \ref{sec:con}.

\section{Discovery}
\label{sec:dis}

The first precise localization of an FRB revealed an unambiguous association with a luminous, persistent synchrotron radio source \citep{2017Natur.541...58C}.  Under the assumption that at least a subset of FRBs should have luminous persistent radio counterparts, \citet{2017ApJ...846...44O} identified 122 potential FRB hosts by identifying radio sources seen toward galaxies. In most cases, the radio sources are active galactic nuclei that reside in the centers of their host galaxies. However, eleven of these sources were designated as potentially interesting as FRB counterparts because they were not located in their host's nucleus. Of those eleven, \src\ was identified as being the most similar to the one known persistent counterpart to an FRB, both in its luminosity and its association with a small, star-forming galaxy.

The VLA Sky Survey started observations in late 2017 and the first epoch (``epoch 1.1'') covers 50\% of the observable sky over the frequency range 2 to 4~GHz to a $1\sigma$\ sensitivity of 120~$\mu$Jy beam$^{-1}$. Pipeline processed images are made available publicly within two weeks of observation. We searched these images for counterparts to the 11 most interesting sources in \citet{2017ApJ...846...44O}. Six of the eleven sources have VLASS coverage in epoch 1.1 and are summarized in Table \ref{tab:ofek}. All of the sources with VLASS coverage have a 3~GHz flux less than that of the 1.4\,GHz flux measured in the FIRST survey. Assuming no variability between FIRST and VLASS, all but one source have implied spectral indexes ($\alpha_{1.4/3~\rm{GHz}}$ for $S_\nu\propto\nu^\alpha$) range from --0.2 to --1.3.

One of the sources in Table \ref{tab:ofek}, \srcfull, is not detected at 3~GHz to a nominal $3\sigma$\ limit of 0.37~mJy. We measured fluxes in VLASS quick-look images with fluxes accurate to 10\% \citep{vlass}, so we estimate a 3~GHz upper limit of 0.4~mJy. The non-detection implies a factor of 50 drop relative to the FIRST survey,\footnote{Spurious variables and transients were found in FIRST data prior to 2012 \citep[see][]{2010ApJ...711..517O}, but those issues have now been corrected \citep{2015ApJ...801...26H}.} equivalent to a factor of 30 drop at fixed frequency (assuming a typical synchrotron spectral index of --0.7). The magnitude of this drop is highly anomalous, and motivated us to search other archival radio data to more fully characterize this source.

\begin{table}[tb]
\caption{VLASS Observations of \citet{2017ApJ...846...44O} Sources}
\centering
\begin{tabular}{lccc}
\hline
Source designation  & $S_{\rm{FIRST}}$\tablenotemark{a} & $S_{\rm{VLASS}}$\tablenotemark{b} & Spectral index \\
in FIRST & (mJy) & (mJy) & \\ \hline
J092758.2--022558 & 2.1 & 1.54 & --0.4 \\
J104726.6+060247 & 2.9 & 1.08 & --1.3 \\
J235351.4+075835 & 4.2 & 3.19 & --0.4 \\
J131441.9+295959 & 2.2 & 1.83 & --0.2 \\
J162244.5+321259. & 2.0 & 1.23 & --0.6 \\
J141918.9+394036 & 20.11 & $<0.4$ & $<-5.1$ \\ \hline
\end{tabular}
\tablenotetext{a}{Typical FIRST peak flux density error is 0.14 mJy.}
\tablenotetext{b}{Typical VLASS peak flux density error in quick-look images is 0.12 mJy (statistical) plus 10\% (systematic).}
\label{tab:ofek}
\end{table} 

\section{Characterization}
\label{sec:cha}

\subsection{Radio Transient}

\begin{figure*}[tb]
 \includegraphics[width=\textwidth]{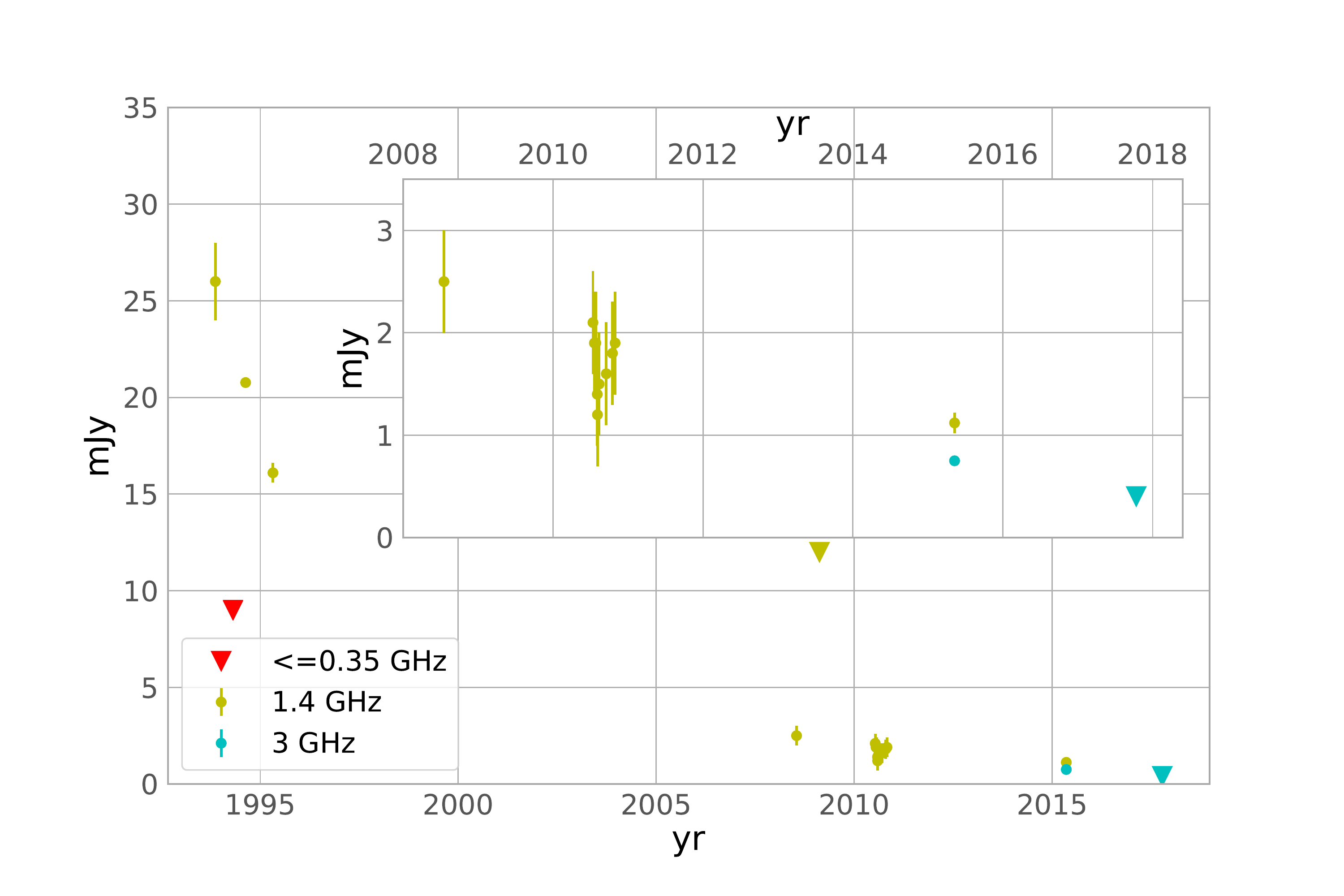}
 \caption{The flux density of \src\ as a function of time, starting with our first detection in 1993. Observing frequencies are shown in three broad ranges ($\le350$~MHz, 1.4~GHz, 3~GHz) encoded with different colors. Upper limits are shown as triangles and detections as circles. Not shown are five upper limits measured between 1977 and 1987 and three of the upper limits at low frequencies determined after 1993. The inset figure shows a closer view of the detections made over the last decade.}
 \label{fig:fluxlc}
\end{figure*}

As shown in Table \ref{tab:obs} and Figure \ref{fig:fluxlc}, the position of \src\ has been observed at 24 other epochs by eight different telescopes over more than four decades. This includes both wide-field sky surveys and serendipitous observations recovered from archives. Below, we describe the analysis of these radio observations.

\begin{table*}[htb]
\caption{Radio Observations of \src. All fluxes have been corrected for primary beam attenuation.}
\centering
\begin{tabular}{lcccc}
\hline
Telescope & Date & Freq. & Peak Flux & Observation Name \\
 & (year) & (GHz) & mJy  & \\ \hline
TI & 1975   & 0.365       & $<400$         & Texas Survey \\
N.\ Cross & 1977 & 0.408 & $<100$ & Bologna Sky Survey \\
NRAO 91m& 1983.3 & 1.4    & $<100$         & GBNSS \\
NRAO 91m& 1987   & 4.85   & $<25$          & GB6 \\
MSRT & 1993.03  & 0.232   & $<180$         & Miyun survey \\
VLA  & 1993.87 & 1.465    & $26\pm2$       & AB6860 \\
VLA  & 1993.87 & 0.325    & $<174$         & AB6860 \\
WSRT & 1994.31 & 0.325    & $<9$           & WENSS survey \\
VLA  & 1994.63 & 1.40     & $20.77\pm0.17$ & FIRST survey  \\
VLA  & 1995.32 & 1.40     & $16.10\pm0.51$ & NVSS survey  \\
VLA  & 2005.20 & 0.074    & $<74$          & VLSSr survey \\
WSRT & 2008.54 & 1.415    & $2.5\pm0.2$    & ATLAS-3D\tablenotemark{a} \\
ATA  & 2009.12 & 1.43     & $<12$          & ATATS \\
WSRT & 2010.53 & 1.415    & $2.1\pm0.2$    & ATLAS-3D \\
WSRT & 2010.55 & 1.415    & $1.9\pm0.2$    & ATLAS-3D \\
WSRT & 2010.57 & 1.415    & $1.9\pm0.2$    & ATLAS-3D \\
WSRT & 2010.59 & 1.415    & $1.4\pm0.2$    & ATLAS-3D \\
WSRT & 2010.59 & 1.415    & $1.2\pm0.2$    & ATLAS-3D \\
WSRT & 2010.61 & 1.415    & $1.5\pm0.2$    & ATLAS-3D \\
WSRT & 2010.71 & 1.415    & $1.6\pm0.2$    & ATLAS-3D \\
WSRT & 2010.79 & 1.415    & $1.8\pm0.2$    & ATLAS-3D \\
WSRT & 2010.82 & 1.415    & $1.9\pm0.2$    & ATLAS-3D \\
GMRT & 2011.29 & 0.15     & $<30$          & TGSS survey \\
VLA  & 2015.36 & 1.52      & $1.1\pm0.1$   & 15A-033 \\
VLA  & 2015.36 & 3.0      & $0.75\pm0.05$  & 15A-033 \\
VLA  & 2017.78 & 3.0      & $<0.4$       & VLASS survey, epoch 1.1 \\ \hline
\end{tabular}
\tablenotetext{a}{The quoted ATLAS-3D flux density errors include an estimate of systematic error added in quadrature to the typical statistical error of 0.1~mJy.}
\tablecomments{Upper limits are as quoted in original work, or 3$\sigma$  when derived from our own analysis.}
\tablerefs{
Texas Survey \citep{1996AJ....111.1945D}, 
Bologna Sky Survey \citep{1985A&AS...59..255F},  
GB6 \citep{1991ApJS...75....1B, 1996ApJS..103..427G},  
GBNSS \citep{1985AJ.....90.2540C, 1992ApJS...79..331W},
Miyun Survey \citep{1997A&AS..121...59Z},  
FIRST \citep{1995ApJ...450..559B},
WENSS \citep{1997A&AS..124..259R},
NVSS \citep{1998AJ....115.1693C},
VLSSr \citep{2014MNRAS.440..327L},
TGSS \citep{2017A&A...598A..78I},
VLASS \citep{vlass},
ATLAS-3D \citep{2012MNRAS.422.1835S},
ATATS \citep{2010ApJ...719...45C}}
\label{tab:obs}
\end{table*} 

\subsubsection{VLA Surveys}

In addition to the FIRST and VLASS sky surveys discussed above, \src\ was observed and detected by the NRAO-VLA Sky Survey at 1.4~GHz \citep[NVSS;][]{1998AJ....115.1693C}. We downloaded images from all three VLA surveys and used {\tt Aegean} \citep{2018PASA...35...11H} to fit a 2d Gaussian to the total intensity counterpart of \src\ in NVSS, designated NVSS~J141918+394036 (see Table \ref{tab:obs}). We also examined the channel-averaged linear polarization data from NVSS, and see no signal in Stokes $Q$ or $U$ down to a 5-$\sigma$ sensitivity of 1.5~mJy, corresponding to an upper limit on the linearly polarized fraction of 9\%.

The most precise localization is measured by FIRST, which finds (RA J2000, Dec J2000) = (14:19:18.855, +39:40:36.0) with an uncertainty of $0\farcs3$ \citep[dominated by systematics at 1/20 of beam size;][]{1997ApJ...475..479W}. We queried the NRAO archive and found that repeated observations toward \src\ \emph{within} the FIRST and NVSS surveys fell within a one week span. A separate analysis of repeated FIRST observations did not reveal \src\ as variable on that timescale \citep{2011ApJ...742...49T}.

\subsubsection{Other Sky Surveys}

In the 1980s, the NRAO 91m telescope was used to conduct sky surveys at 1.4 and 4.85~GHz \citep{1985AJ.....90.2540C,1991ApJS...75....1B}. These surveys are interesting because they were relatively sensitive and observed only a few years before the first VLA detection of \src\ in the 1990s. \src\ was not detected in these surveys to limiting flux densities of 100 and 25~mJy at 1.4 and 4.85~GHz, respectively.

Six other sky surveys observed the position of \src\ at frequencies below 1~GHz: the VLA at 74 MHz (VLSSr), the GMRT at 150 MHz (TGSS), the MSRT at 232 MHz (Miyun Survey), WSRT at 325 MHz (WENSS), the Texas Interferometer at 365 MHz (Texas Survey), and the Northern Cross at 408~MHz (Bologna Sky Survey). None of these surveys detected the transient (see Table~\ref{tab:obs}). 

WENSS is not only the most sensitive of the low-frequency surveys, but it observed this field within a few months of VLA observations near the (apparent) peak brightness at 1.4~GHz. Assuming a months-scale evolution in the \src\ radio brightness, the WSRT and FIRST observations are effectively simultaneous and the WSRT non-detection limits the spectral index near peak flux, $\alpha_{0.35/1.4}>+0.6$. 

The Allen Telescope Array (ATA) observed \src\ as part of a 690 deg$^2$\ region at 1.4~GHz \citep[the ATATS survey;][]{2010ApJ...719...45C}. No counterpart was detected above its $3\sigma$\ flux density upper limit of 12~mJy. 

\subsubsection{WSRT Archives}
\label{sec:wsrt}

A search of the WSRT archives revealed a series of observations of NGC 5582 (roughly 10\arcmin\ away) as part of the ATLAS-3D project \citep{2012MNRAS.422.1835S}. In total, ten epochs of observing were made in 2008 and 2010 with a bandwidth of 20~MHz centered at 1.415~GHz (Oosterloo, Morganti and Serra, in prep). The data, which are in total intensity only, were calibrated and corrected for primary-beam attenuation with {\tt MIRIAD} \citep{1995ASPC...77..433S}, and images at each epoch detect \src\ with a significance of roughly 20$\sigma$. 

We modeled the source in FITS images with {\tt Aegean} and quote peak flux densities in Table \ref{tab:obs}. The statistical errors are roughly 0.1~mJy, but a joint analysis of all sources in the field suggests that there are significant systematic effects also. We find that the standard deviation of peak flux for all sources scales with brightness and has a minimum of 0.2~mJy. We add this value to the statistical error in quadrature for all analysis presented here.

The WSRT fluxes for \src\ in Table~\ref{tab:obs} do not show a purely secular decay, but instead suggest some small level of gradual variability up and down across epochs. To assess the possibility of short-term flux variability, we identified 22 other unresolved sources within the WSRT field of view, each detected at all or almost all epochs, and ranging in flux from 1 to 380~mJy. For each source, we extracted fluxes at each epoch, and computed the mean flux $\mu$, standard deviation $\sigma$, and modulation index $m \equiv \sigma/\mu$. 
\src\ has $m = 0.21$, which is larger than for any of the other 22 sources, for which the mean value is $m = 0.10\pm0.05$. However, the first flux measurement was made two years prior to the others, so the apparent long-term decay of the flux of \src\ increases the modulation index. If we disregard the flux point from 1998, we find $m = 0.17$ for \src, which is comparable to that for other millijansky sources in the field: for example, FIRST J142131.9+392355 has a mean WSRT flux of 1.5~mJy with $m = 0.20$ and a randomly jittering flux across the WSRT epochs, but has a somewhat different FIRST flux of 2.3 mJy; on the other hand,  FIRST~J142103.2+392448 has $\mu = 2.2$~mJy and $m = 0.15$ with a smoothly varying light curve like seen for \src, but has a similar FIRST flux of 2.3 mJy. Overall, we conclude that there is the suggestive possibility from the data that \src\ exhibits flux variability, but we cannot definitively confirm or rule out such behavior from the WSRT data.

\subsubsection{VLA Archives}

A search of the VLA archives revealed two VLA observing campaigns with useful data toward \src. The first of these is a legacy VLA observation AB6860, which observed in Nov~1993 (about nine months earlier than FIRST), at both 325~MHz and 1.4~GHz.

We analyzed the AB6860 observations in {\tt MIRIAD} and detected the source with a significance of $40\sigma$ at 1.4~GHz (the 325-MHz observations result only in an upper limit). The AB6860 observations did not contain a flux calibrator, so the flux scale is estimated via considering the fluxes of six field sources, all detected at unresolved objects in the FIRST, NVSS and ten WRST observations described above. Two of these sources (FIRST J141858.8+394626 and FIRST J142006.4+393503) show significant ($>$50\%) changes in flux between epochs, and are assumed to be variable sources. The remaining four sources (FIRST J142009.3+392738, FIRST J142030.5+400333, FIRST J142120.6+394110 and FIRST J142123.5+393332) have consistent fluxes at 1.4~GHz across all 12 epochs of approximately 35, 12, 15 and 380~mJy, respectively. Setting the fluxes of these four sources to these values in the AB6860 data, we determine a 1.4-GHz flux for \src\ at epoch 1993.87 of 27~mJy. We note though that there is a 4\%--5\% difference in observing frequency between the AB6860 observations and these other data (see Table~\ref{tab:obs}). 
Assuming a typical radio galaxy spectral index for these four calibration sources of --0.7, and using the scatter between the flux measurements from FIRST, NVSS and WSRT as an estimate of the uncertainties, we determine a 1.4-GHz flux for \src\ at this epoch of $26\pm2$~mJy as listed in Table \ref{tab:obs}. We see no linearly or circularly polarized emission from \src\ in these data above a 5$\sigma$ limit of $\sim4$~mJy, corresponding to a fractional upper limit of $\sim15$\%.

The second archival VLA observation was conducted at 1.4 and 3.0~GHz in 2015 (a ``Jansky'' VLA project designated 15A-033\footnote{In a remarkable coincidence, one of us (B.M.G.) was a co-investigator on this archival observation.}). We calibrated and imaged this observation using CASA \citep{2007ASPC..376..127M}. \src\ is detected in both the 1.4 and 3~GHz bands at a brightness of about 1~mJy with a significance of 10-15$\sigma$. The mean flux across the 1.4 and 3~GHz bands implies a spectral index of $\alpha_{1.4/3}=-0.6\pm0.2$. This late-time spectral index measurement has changed significantly from the early lower limits on the spectral index between 0.35 and 1.4~GHz.

\subsection{Association with a Host Galaxy}
\label{sec:host}

\begin{figure}[tb]
 \includegraphics[width=\columnwidth]{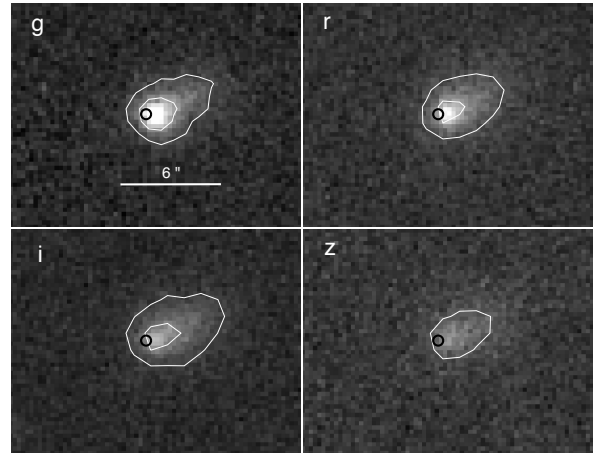}
  \caption{Comparison of \src\ and its host galaxy as seen by Pan-STARRS in $g$, $r$, $i$, and $z$\ bands. Each panel is centered on the stacked position of SDSS J141918.81+394035.8 (a.k.a.~ PSO J141918.804+394035.996) with contours at 3 and 10 times the noise in each image. The FIRST radio localization of \src\ is shown as a black circle with radius 0\farcs3.}
  \label{fig:host}
\end{figure}

\src\ is located $0\farcs5$ from the center of the galaxy SDSS J141918.81+394035.8 \citep{2018ApJS..235...42A} with an r-band magnitude $\sim19$. The galaxy is also detected by Pan-STARRS \citep{2016arXiv161205560C}, the Dark Energy Camera Legacy Survey \citep[DECaLS;][]{2018arXiv180408657D}, the intermediate Palomar Transient Factory  \citep[iPTF;][]{2017PASP..129a4002M}, and the USNO-B survey \citep{2003AJ....125..984M}. Figure \ref{fig:host} shows the association of \src\ with the galaxy. The iPTF DR3 catalog shows it was detected in 34 epochs spanning 2.3 years around 2009 and had no significant variability. The USNO-B1 catalog shows that the galaxy had a similar magnitude in 1979.

An SDSS spectrum of the galaxy\footnote{See \url{https://dr9.sdss.org/spectrumDetail?plateid=1380&mjd=53084&fiber=534}} finds strong emission lines indicative of active star formation. These lines provide a robust redshift measurement of $z=0.01957$ \citep[a~distance of 87~Mpc using $H_0 = 67.4$~km~s$^{-1}$~Mpc$^{-1}$;][]{2018arXiv180706209P}. At this distance, the galaxy has an extinction-corrected absolute r-band magnitude of --16 and a size of about 2.5~kpc (6\arcsec\ in projection). The $g$- and $r$-band optical images show a clear enhancement on the east side of the galaxy that is coincident with \src. Dwarf galaxies such as this one can host supermassive black holes \citep[e.g., Henize 2-10;][]{2016ApJ...830L..35R}, but an enhancement at the galaxy's edge is more likely to be an area of enhanced star formation.

SDSS J141918.81+394035.8 is also detected by WISE,\footnote{The WISE emission from SDSS~J141918.81+394035.8 plus the FIRST detection of \src\ gave the galaxy an apparent radio to infrared flux ratio otherwise seen only for radio-loud active galactic nuclei at redshifts $z>2$ \citep{2014MNRAS.439..545C}.} which allowed \citet{2015ApJS..219....8C} to combine with SDSS data to model this galaxy's star formation properties. They find that SDSS J141918.81+394035.8 has a stellar mass $\log{M/M_\odot}=7.24\pm0.11$, a star formation rate  $\log{\rm{SFR}/(M_\odot yr^{-1})}=-1.2^{+0.16}_{-0.2}$, and specific star formation rate $\log{\rm{sSFR}/yr^{-1}}=-8.4\pm0.2$  (95\% confidence intervals). The low mass and high specific star formation rate make this galaxy similar to the typical host galaxies of long GRBs and superluminous supernovae \citep{2016A&A...593A.115J}.

The FIRST catalog includes about $9.5\times10^5$\ sources brighter than 1~mJy in an area of $1.1\times10^4$ deg$^{2}$.  The probability of a single FIRST source lying within 1\arcsec\ of a host galaxy is about $2\times10^{-5}$. However, the \citet{2017ApJ...846...44O} sample was created by associating a large sample of galaxies with a FIRST source with multiple additional requirements (e.g., compact radio source, a good quality radio flux). Considering the combined size of all compact galaxies under these selection criteria, we calculate a chance of random association of $\sim0.1$. Although variability was not a selection criterion, we note that variable FIRST sources are 10$^3$\ times less common.\footnote{In fact, \src\ is among the most variable FIRST sources known \citep{2006ApJ...639..331G,2011ApJ...742...49T}.} Furthermore, \src\ is coincident with a star-forming region within the galaxy, which is independent confirmation of association under any source model that scales with star formation. Therefore, from here forward we assume that \src\ is associated with SDSS J141918.81+394035.8.  However, if this is not the case and if \src\ is instead a background extragalactic source, we note that the energetic requirements on the transient would further increase. 

The association of \src\ with the host allows us to measure a peak spectral luminosity of $2\times 10^{29}$\ erg s$^{-1}$\ Hz$^{-1}$ at 1.4~GHz and $\nu L_\nu \gtrsim 3\times10^{38}$~erg s$^{-1}$. Since we do not detect a rising part of the radio light curve, this is a lower limit on the peak luminosity.

\subsection{Search for a Gamma-Ray Counterpart}
\label{sec:GRB}

We have examined the 4Br BATSE gamma-ray burst catalog \citep{1999ApJS..122..465P} and NASA’s ``GRBCAT'' compilation\footnote{This also includes GRBs detected by other missions and by the interplanetary network (IPN); see \url{https://heasarc.gsfc.nasa.gov/grbcat/}} for possible GRBs coincident with \src\ and occurring prior to our earliest radio detection of \src\ in Nov~1993. We find one burst, GRB 921220, whose error circle encompasses our position for \src. However, the positional uncertainty for this burst is enormous, covering more than 1500~deg$^2$ ($\sim$4\% of the entire sky), and we thus do not find this coincidence significant. The nearest other burst to \src\ was GRB 911024, whose nominal position was 5.0 degrees from \src, and for which the statistical positional uncertainty is a circle of radius 4.13 degrees. The full positional uncertainty also includes a systematic term \citep[see][]{1999ApJS..122..503B}, which may make the position of \src\ marginally consistent with GRB 911024. However, given that there are $\sim1000$ known GRBs occurring prior to Nov~1993, the chances of any random position on the sky falling within a few degrees of an unrelated GRB is close to unity. We therefore cannot identify any convincing candidates for a GRB associated with \src.

The lack of any obvious gamma-ray counterpart by BATSE and the IPN is very constraining for nearby GRBs \citep{2013ApJ...769..130C}. From 1991 to 1993, Ulysses was part of the IPN and it had an largely unrestricted view of the sky. Ulysses was essentially flux complete to BATSE GRBs above $10^{-4}$~erg~cm$^{-2}$ \citep{1999ApJS..122..503B, 1999ApJS..120..399H}. Most GRBs have an isotropic energy greater than $10^{51}$~erg in gamma rays \citep{2002A&A...390...81A, 2009ApJ...703.1696Z}, which corresponds to a fluence of $10^{-3}$~erg cm$^{-2}$ at a distance of 87~Mpc. The non-detection by both BATSE and the IPN effectively rules out the presence of an on-axis GRB.  It would not, however, necessarily rule out the presence of less-energetic (``low luminosity") GRBs, which have isotropic energies $\sim 10^{48}-10^{50}$ erg (e.g.,~\citealt{2007ApJ...654..385K}) and have been argued on theoretical grounds to originate from a distinct emission mechanism, such as shock break-out, that is more isotropic than the jetted GRB emission (e.g.,~\citealt{2012ApJ...747...88N}).

\subsection{Rates}
\label{sec:rates}

\src\ is the only non-nuclear FIRST source found to be so strongly variable (Table \ref{tab:ofek}) in comparison to VLASS and here we estimate the volumetric rate of \src-like transients. While the study of \citet{2017ApJ...846...44O} is complete above 1.5~mJy, we estimate that \src\ would only have been identified if it were brighter than roughly 4~mJy at peak, since this would require an unphysically steep spectral index when compared between 1.4 and 3~GHz. From Figure~\ref{fig:fluxlc}, we see that the source exceeded 4~mJy for a window of $\Delta t \sim 2000$ days. 

The radio catalog was drawn from the FIRST survey, which covered $\Delta \Omega\sim10,600$ deg$^{2}$\ or 25\% of the sky. Since VLASS epoch 1.1 covers half the visible sky and about half of the \citet{2017ApJ...846...44O} catalog, we estimate its completeness relative to FIRST as $f_{\rm{VLASS}}\sim0.5$. \citet{2017ApJ...846...44O} estimated the completeness of his spectroscopic galaxy sample as roughly $f_{\rm{spec}}\sim30$\% of the stellar luminosity out to distance of $d\sim108$~Mpc. If we assume that the event rate scales with stellar luminosity, we find a volumetric rate of
\begin{equation}
\mathcal{R} \approx \frac{1\rm \,event}{\Delta t(f_{\rm{spec}} f_{\rm{VLASS}} \Delta \Omega d^{3}/3)} \approx 900\,{\rm Gpc^{-3}yr^{-1}}.
\end{equation}
This rate is only accurate to an order of magnitude due to uncertainties in estimating its time scale, galaxy selection, and human bias. We conservatively define the rate constraint as a 95\% confidence interval on a Poisson rate of 50--4200 Gpc$^{-3}$~yr$^{-1}$ \citep{1986ApJ...303..336G}.

For comparison, the rate of on-axis long GRBs (LGRBs) with $E_{\rm K} \gtrsim 10^{51}$ ergs is $\mathcal{R}_{\rm LGRB}(z = 0) \approx 0.3$ Gpc$^{-1}$ yr$^{-1}$ \citep{2005ApJ...619..412G}. For a beaming-correction of $f_{\rm b}^{-1} = 200$ \citep{2016ApJ...818...18G}, we expect a true off-axis LGRB rate of $\mathcal{R}_{\rm off-axis\,LGRB} \approx 60\,{\rm Gpc^{-3}yr^{-1}}$. This rate is consistent with the estimated rate for \src-like transients.

\subsection{Summary}

\src\ is a radio transient characterized by:
\begin{itemize}
 \item association with a star-forming region in a dwarf galaxy,
 \item a peak transient luminosity at 1.4~GHz of $2\times10^{29}$~erg s$^{-1}$ Hz$^{-1}$ or $\nu L_\nu \gtrsim 3\times10^{38}$~erg s$^{-1}$,
 \item a radio peak lasting $>1.5$~yr and detections spanning 21.5 years,
 \item a radio spectral index ($S_\nu\propto\nu^\alpha$) that evolves from $> +0.6$\ at peak to $\sim -0.6$ at late times,
 \item no associated gamma-ray burst, and
 \item an inferred rate between 50 and 4200 Gpc$^{-3}$~yr$^{-1}$ (95\% confidence interval).
\end{itemize}
The location of the transient at the edge of its host argues that it is not associated with some exceptional flare from an active galactic nucleus. Extreme scattering events and other radio propagation-related phenomena have not been observed to magnify to the degree seen for \src\ \citep{2016Sci...351..354B}, so we conclude that it is likely an energetic transient. 

The peak detected radio luminosity of \src\ exceeds that of supernovae associated with low circumburst densities, such as SN Ibc \citep[$<10^{28}$ erg s$^{-1}$ Hz$^{-1}$ on timescales of 1 to 30 years;][]{2006ApJ...638..930S, 2010Natur.463..513S}, SN Ia \citep[$<10^{25}$~erg s$^{-1}$~Hz$^{-1}$;][]{2012ApJ...750..164C}, and short GRBs \citep[$<10^{29}$ erg s$^{-1}$ Hz$^{-1}$ on timescales of years;][]{2014MNRAS.437.1821M, 2016ApJ...831..141F}. The long timescale of \src\ is similar to that observed for supernovae in dense environments \citep[e.g., SN IIp and IIn;][]{1991ApJ...380..161W, 1996AJ....111.1271V, 2015ApJ...810...32C}, but those systems are less luminous. 

The limits on radio timescale and luminosity of \src\ are consistent with observations of long GRBs \citep{2002ARA&A..40..387W} and TDEs \citep{2012ApJ...748...36B, 2016ApJ...819L..25A}, and with theoretical predictions for the class of transients powered by magnetar spin-down \citep[e.g.,][]{2015MNRAS.454.3311M, 2016MNRAS.461.1498M, 2017ApJ...841...14M, 2018arXiv180605690M}. The host galaxy is similar to those known to host long GRBs and superluminous supernovae \citep{2006Natur.441..463F,2015ApJ...804...90L}.

Finally, we note that there may be two radio transients with measured luminosities that are potential analogs to \src. J060938--333508 was a transient detected in a single epoch of a 843~MHz survey of the southern sky \citep{2011MNRAS.412..634B}. It is seen in projection on the disk (non-nuclear) of a galaxy at $z=0.037$, implying a luminosity $6\times10^{29}$~erg cm$^{-2}$~s$^{-1}$ with an unconstrained time scale. Second, \citet{1994ApJ...420..152Y} discovered Markarian 279A, a radio transient with peak luminosity $1\times10^{29}$~erg s$^{-1}$~Hz$^{-1}$ that dropped to half peak after $\sim$2 years at 1.4~GHz. No optical or high-energy transient has been associated with either of these events.

\section{Transient Origin}
\label{sec:ori}

In this section we consider possible origins of \src.  We begin with most commonly considered type of extragalactic radio transients, namely that created as the high-velocity ejecta from an explosive event, such as a GRB, interacts with its gaseous external environment.  Then, we explore an alternative hypothesis that \src\ is the young wind nebula of a compact object, such as a flaring magnetar, embedded within a slower-expanding supernova ejecta shell.  

\subsection{Synchrotron Blast Waves}

\subsubsection{General Considerations}

 \citet{2015ApJ...806..224M} derive general constraints on the light curves of synchrotron radio transients from explosions interacting with a gaseous external medium. They consider the ejection of material with a kinetic energy of $E_K= 10^{51}\,E_{K,51}$\ erg and an initial velocity of $v_i=\beta_i c $ (corresponding to an initial Lorentz factor of $\Gamma_i= [1-\beta_i^{2}]^{-1/2}$), into an ambient medium of constant density $n=1\,n_{0}$ cm$^{-3}$. The ejecta transfer their energy to the ambient medium on a deceleration timescale:
\begin{equation}
  t_{\rm dec} \approx R_{\rm dec}/2c\beta_i\Gamma_i^{2} \approx
  115\,{\rm d}\,\,E_{K,51}^{1/3}n_{0}^{-1/3}\beta_i^{-5/3}\Gamma_i^{-8/3}.
\label{eq:tdec}
\end{equation} 
where $R_{\rm dec}$ is the characteristic radius at which the ejecta have swept up a mass $\sim 1/\Gamma_i$ of their mass.  The timescale $t_{\rm dec}$ defines a lower limit on the rise time to peak for radio transients (see below).

If the observing frequency ($\nu_{\rm obs} = 1.4\,\nu_{1.4}$ GHz) is located above both the synchrotron peak frequency ($\nu_m$) and the self-absorption frequency ($\nu_a$), then the peak brightness is achieved at $t_{\rm dec}$, and is given by \citep{2011Natur.478...82N},
\begin{equation}
  F_{\nu,\rm dec} \approx 70\,{\rm mJy}\,\, E_{K,51} n_{0}^{4/5}
  \epsilon_{e,-1}^{6/5}\epsilon_{B,-2}^{4/5}\beta_i^{9/4}\nu_{\rm 1.4}^{-0.6},
\label{eq:Fp}
\end{equation} 
where we have scaled the distance to that of \src.  We make the standard assumption that electrons are accelerated at the shock front into a power-law distribution in momentum with slope $p$ (as predicted by the theory of Fermi acceleration at shocks; \citealt{blandford_eichler_87}) and that $\epsilon_B = 0.01\,\epsilon_{B,-2}$\ and $\epsilon_e= 0.1\, \epsilon_{e,-1}$\ are the fractions of post-shock energy in the magnetic field and relativistic electrons, respectively.  We have taken $p = 2.2$ for consistency with the late-time 1.4/3~GHz spectral index of \src, assuming it arises from optically-thin synchrotron emission with $F_{\nu} \propto \nu^{-(p-1)/2}$.  The rise time can be longer than $t_{\rm obs}$, and the peak flux somewhat lower, at observing frequencies below the self-absorption frequency 
\begin{equation} \nu_{\rm sa}(t_{\rm dec}) \approx 0.75{\rm GHz}\,E_{K,51}^{0.11}n_{0}^{0.55}\epsilon_{e,-1}^{0.39}\epsilon_{B,-2}^{0.34}\beta_{i}^{1.23}
\label{eq:nusa}
\end{equation}

Relativistic transients, such as GRBs and jetted TDEs, produce ultra-relativistic ejecta in tightly collimated jets.  Less relativistic transients produce essentially spherical ejecta with isotropic emission at all times. However, relativistic transients viewed in an initial off-axis direction can also be approximated as being spherically symmetric once the shocked matter decelerates to sub-relativistic velocities and spreads laterally into the observer's line of sight \citep[e.g.,][]{2009ApJ...698.1261Z, 2011ApJ...738L..23W}.  At this point the radio emission is no longer strongly beamed, and $t_{\rm dec}$ and $F_{\nu,\rm dec}$ can be approximated using a mildly-relativistic value $\beta_i \approx 0.5$.  

We focus our analysis of \src\ on the case of non-relativistic ejecta, or an initially relativistic but off-axis jet.  At the relatively low observing frequencies of interest, an off-axis viewing angle is the most likely configuration for a volume- or flux-limited sample of radio transients \citep{2015ApJ...806..224M}.  From Equation~(\ref{eq:Fp}), we then see that for the fiducial values of the microphysical parameters and an external density $n_0 \sim 1-10$ cm$^{-1}$ typical of the average interstellar medium (ISM) of a star-forming galaxy or the circumburst environment of a massive star, explaining the peak 1.4-GHz flux $\gtrsim 20$ mJy for \src\ requires an initially relativistic explosion ($\beta_i \approx 0.5$) with a kinetic energy $E_{\rm K} \sim 3 \times 10^{50}-3\times 10^{51}$ erg.  We can also place an absolute lower limit of $E_{\rm K} \sim 10^{49}$ erg for the case of equipartition $(\epsilon_e = \epsilon_B \sim 0.5$).  For such parameters, the predicted rise-time for an off-axis event is $t_{\rm dec} \sim 0.1-1$ yr (Equation~[\ref{eq:tdec}]).  The predicted self-absorption frequency $\nu_{\rm sa}(t_{\rm dec}) \gtrsim 0.3-3$ GHz (Equation~[\ref{eq:nusa}]), below which one predicts a spectral turnover to $F_{\nu}\propto \nu^{2}$, is also consistent with the early-time non-detection of \src\ at $\nu \le $ 350 MHz.  The fact that the self-absorption frequency is close to the 1.4 GHz band at the discovery epoch around 1994 also suggests the transient was detected close to its peak; although a nice consistency check, this makes the earlier archival upper limits (e.g.,~$< 25$ mJy at 4.85~GHz in 1987) relatively unconstraining. 

Alternatively, the peak flux could be consistent with a spherical non-relativistic explosion $\beta_i \ll 1$ with a larger energy $E_{\rm K} \sim 4\times 10^{51}{\rm erg} \,n_{0}^{-4/5}(\beta_i/0.3)^{-9/4}$.  An example of such a transient is the ejecta from a neutron star merger ($\beta_i \sim 0.1-0.3$), possibly energized by a long-lived magnetar remnant \citep{2014MNRAS.437.1821M}.  The rate of neutron star mergers of $1540^{+3200}_{-1220}$ Gpc$^{-3}$ yr$^{-1}$ inferred from the discovery of GW170817 (\citealt{LIGO+17DISCOVERY}) agrees with that of \src\ (see \S\ref{sec:rates}).  However, one reason to disfavor this scenario as compared to the long GRB case is the the small size of the star-forming host of \src; the lowest measured host galaxy mass of a short GRB is $10^{8.8}M_{\odot}$ \citep{2010ApJ...725.1202L}, an order of magnitude larger than the host of \src.  

\subsubsection{GRB Afterglow}
\label{sec:afterglow}

Based on arguments given above, the peak flux and characteristic timescale of \src\ are broadly consistent with arising from a blast wave of energy and external density similar to those which characterize long-duration GRB jets. The long GRB scenario also agrees with the otherwise peculiar host galaxy properties of \src\ and, albeit marginally, estimates of the occurrence rate for off-axis events (\S\ref{sec:rates}). 

\begin{figure}[tb]
 \includegraphics[width=\columnwidth]{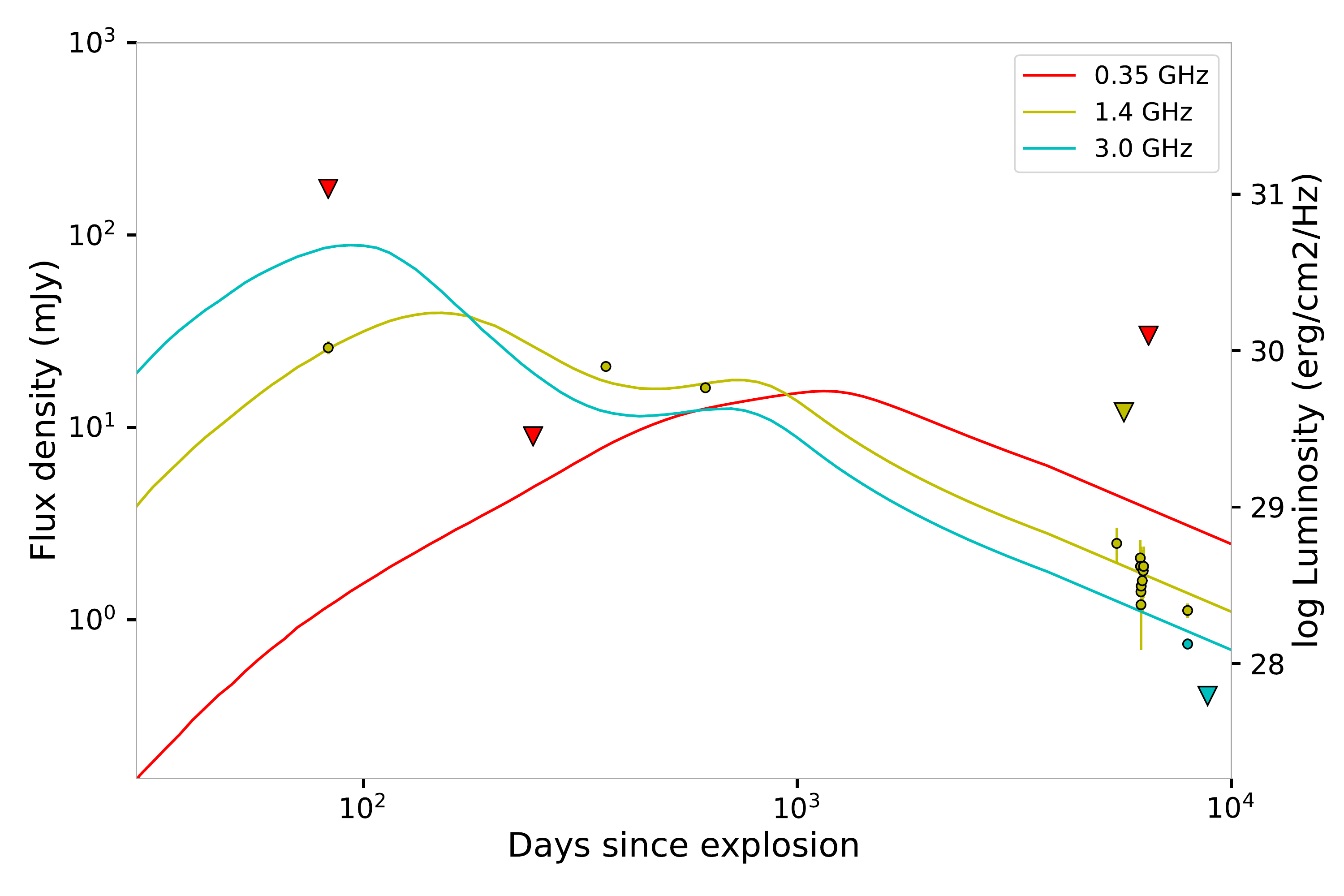}

 \includegraphics[width=\columnwidth]{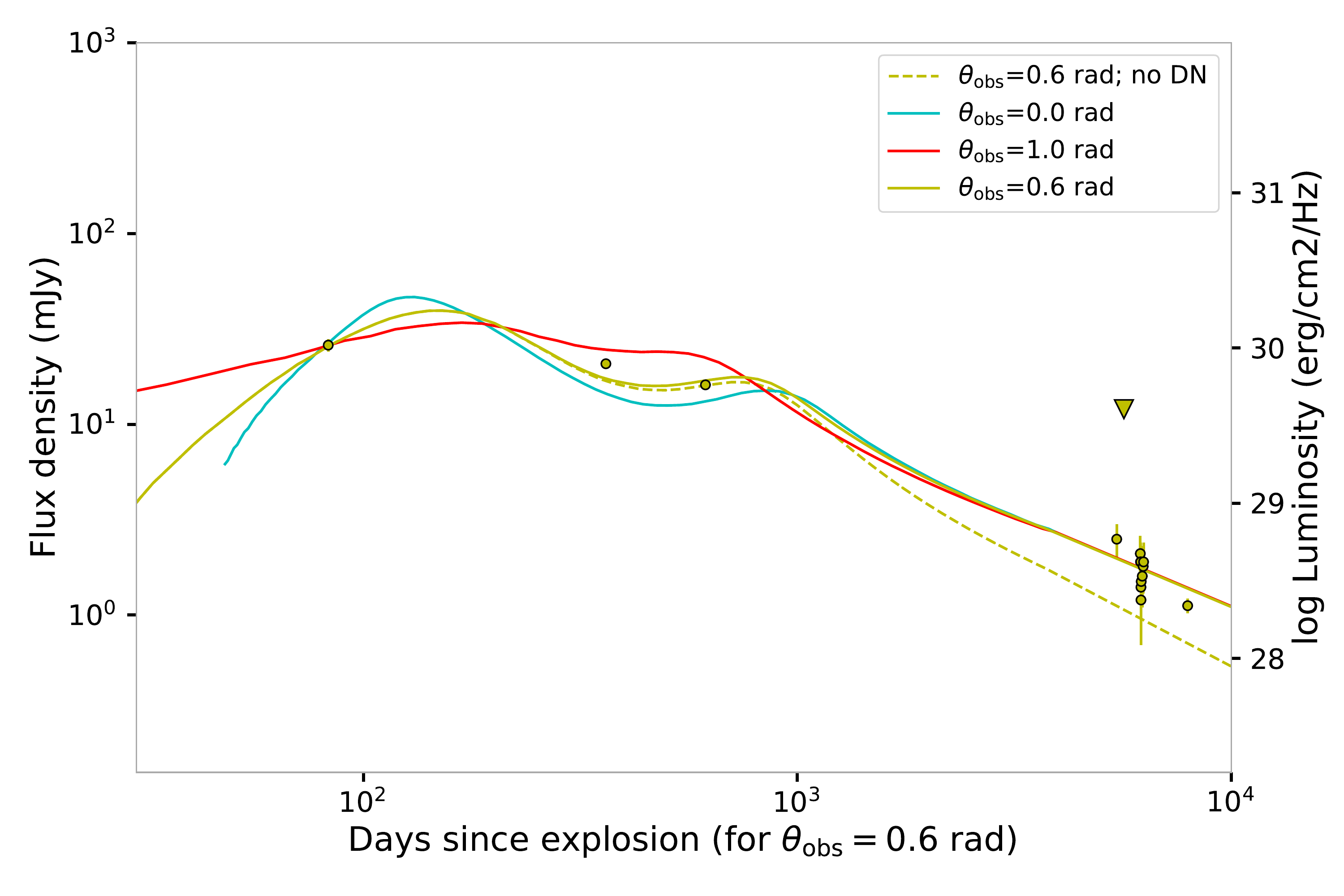}
  \caption{Radio light curve of \src\ starting in 1993 compared to GRB afterglow models that include deep Newtonian corrections \citep{2012ApJ...749...44V, 2013ApJ...778..107S}. (Top) The measurements are shown with red, yellow, and cyan representing 0.35, 1.4, and 3.0~GHz fluxes, respectively. The lines use the same color scheme and show the model fluxes for a GRB viewing angle $\theta_{\rm{obs}}=0.6$~rad with explosion date 83 days before the first measurement in late 1993. The isotropic energy is $E_{\rm iso} \approx 2\times 10^{53}$ erg (giving a beaming-corrected energy of $E_K\sim 10^{51}$ ergs), constant ISM density $n = 10$ cm$^{-3}$, electron index $p = 2.2$, and microphysical parameters $\epsilon_e = 0.1$, $\epsilon_{B} = 0.025$. (Bottom) A set of afterglow models is compared to measurements at 1.4~GHz. The $\theta_{\rm{obs}}=0.6$~rad model is the same as in the top panel and defines the x-axis labels. The dashed line shows 1.4~GHz predictions without corrections for the deep Newtonian phase. The cyan and red lines show how the 1.4~GHz predictions change for viewing angles of 0.0 (on axis) and 1.0~rad, respectively. The explosion dates are set to fit to the first 1.4~GHz measurement and occur 45 and 180 days before that measurement, respectively.
  \label{fig:model}}
\end{figure}

To explore this connection in greater depth, Figure~\ref{fig:model} compares the multi-band light curves of \src\ to theoretical predictions for GRB afterglows based on relativistic hydrodynamical simulations of a relativistic jet interacting with a constant density ISM from the afterglow library models of \citet{2012ApJ...749...44V}, modified as described by \citet{2013ApJ...778..107S} (see below).  We explored a range of parameters and tested for consistency with the peak luminosity, timescale of evolution, early-time spectral index, and late-time decay.  We find reasonable agreement between radio measurements at 0.35, 1.4, and 3.0~GHz and models for assumed microphysical parameters $\epsilon_e = 0.1$, $\epsilon_{B} = 0.025$, electron spectral index $p = 2.2$, external density $n = 10$ cm$^{-3}$, and isotropic jet energy $E_{\rm iso} = 2\times 10^{53}$ erg, the latter corresponding to a beaming-corrected jet energy of $E_{\rm K} \approx 10^{51}$ erg.  All of these parameters are well within the range of those inferred by modeling the afterglows of on-axis cosmological long GRBs (e.g.,~\citealt{2015ApJ...799....3R}).  The implied time of the burst from these models ranges from a few months to a year prior to the first VLA epoch in late 1993.  

The observer viewing angle relative to the jet axis, $\theta_{\rm obs}$, is not well-constrained by the data, as its value is somewhat degenerate with the assumed microphysical parameters; our ``best fit" model shown in the top panel of Figure~\ref{fig:model} has $\theta_{\rm obs} \approx 0.6$, but as shown in the bottom panel, a range of other values $\theta_{\rm obs} \lesssim 1.0$ can also work given the uncertainties.  The stronger evidence for an off-axis viewing angle comes from the lack of a coincident GRB (\S\ref{sec:GRB}), and from the overall much higher geometrical probability of seeing an off-axis event $\propto 1-\cos \theta_{\rm obs}$.

One exception to the generally good afterglow fits is the most recent 3 GHz non-detection by VLASS, which requires a $\sim 50\%$ drop in flux over a baseline of just two years from the prior detection.  Such a rapid change is challenging to explain in afterglow models; even large discontinuities in the ISM density produce at most order-unity changes in flux \citep{2011MNRAS.418..583M}, but only over timescales comparable to the source age, which in this case is $\gtrsim 23$ years. Alternatively, this flux drop may be a sign of scintillation-induced fluctuation. Reanalysis of VLASS data or new observations are needed to confirm the flux drop.

If \src\ is indeed a GRB afterglow, its nearby distance and long observational baseline as compared to cosmological GRBs enables an unprecedented test of the late-time behavior of afterglow light curves. The standard prediction for late time decay is \citep{2000ApJ...537..191F, 2004ApJ...600..828F}:
\begin{equation}
F_{\nu} \propto t^{-3(5p-7)/10} \underset{p = 2.2}\approx t^{-1.2}
\end{equation}
where we have again used $p = 2.2$ for consistency with the late-time 1.4/3~GHz spectral index. However, these models do not generally include the late-time {\it flattening} of the radio light curve which is expected as the blast wave enters the so-called ``deep Newtonian" regime \citep{2013ApJ...778..107S}. In the deep Newtonian regime, which happens after a timescale $t_{\rm DN} \approx 2.1\,\,{\rm yr}\,E_{\rm K,51}^{1/3}n_{0}^{-1/3}\epsilon_{e,-1}^{5/6}$, the bulk of the shock-accelerated electrons turn non-relativistic and the theory of Fermi acceleration at shocks \citep{blandford_eichler_87} predicts that the electron spectrum should be a power-law distribution in {\it momentum} rather than energy. In that case, the light curve is predicted to decay as a shallower power-law given by 
\begin{equation}
F_{\nu} \propto t^{-3(p+1)/10} \underset{p = 2.2}\approx t^{-0.96},
\label{eqn:newt}
\end{equation}
where $p$\ is the electron distribution power-law slope. 

The bottom panel of Figure~\ref{fig:model} shows that, if we were to neglect the physically-motivated deep Newtonian correction, our model would underpredict the late-time 1.4~GHz flux measurements by nearly a factor of 2. Our observations of \src\ may thus provide some of the first evidence that GRB afterglows decay at late-times in a way consistent with predictions for the deep Newtonian regime \citep{2013ApJ...778..107S}. 

\subsection{Magnetar Birth Nebula}

Our discussion thus far has focused on radio transients from an external blast wave interacting with surrounding gas. Alternatively, \src\ could be fading emission from the wind nebula powered by a young compact object, such as a flaring magnetar, embedded behind the ejecta of a decades-old supernova remnant \citep{2014MNRAS.437..703M, 2016MNRAS.461.1498M,2017ApJ...841...14M, 2018MNRAS.474..573O,2018arXiv180605690M}. While these transients are luminous, their long evolution timescale has made it hard to constrain this population. Indeed, such a model was proposed in the context of the quiescent radio source discovered to be spatially coincident with the repeating fast radio burst FRB 121102 \citep{2017Natur.541...58C,2017ApJ...841...14M}, the search for which was the original motivation of \citet{2017ApJ...846...44O} that led to the identification of \src. 

In this model, the 1.4~GHz radio emission is obscured by free-free absorption through the supernova ejecta shell for the first couple of decades of its evolution (e.g.~\citealt{2016MNRAS.458L..19C,2016ApJ...824L..32P,2018arXiv180605690M}), so its explosion would be significantly earlier than implied in the GRB afterglow model. The non-detection at lower frequencies $\nu \le 350$ Hz around the time of the 1.4 GHz peak flux (just after the ejecta shell has become transparent at this frequency) is also consistent with the $\nu^{-2}$ dependence of the free-free opacity. 

The evolution and spectrum of the radio light curve in this model is uncertain, as it depends on poorly understood details of the rate at which magnetic fields and electrons are injected into the nebula (e.g.~\citealt{2018arXiv180605690M}). Nevertheless, the present-day power-law spectrum of \src\ between 1.4 and 3~GHz is consistent with that measured at low frequencies for the quiescent source of FRB 121102 \citep{2017Natur.541...58C}. If the majority of FRBs share the same luminosity function as FRB 121102, the formation rate of FRB-producing objects is also consistent, albeit with large uncertainties, with that of long-duration GRBs and super-luminous supernovae\footnote{Superluminous supernovae have a rate of $\approx30$~Gpc$^{-3}$~yr$^{-1}$ in the local universe \citep{2017MNRAS.464.3568P}} for an assumed burster lifetime of decades to centuries \citep{2016MNRAS.461L.122L,2017ApJ...843...84N,2017ApJ...850...76L}. Based on rate considerations, \src\ therefore also seems consistent with this model. This model predicts that \src\ is a source of FRB emission and would be detectable if beamed in our direction. Future monitoring of this source is strongly encouraged. 

\section{Conclusions}
\label{sec:con}

We have discovered a decades-long, luminous radio transient, \srcfull, likely hosted by the dwarf galaxy SDSS~J141918.81+394035.8 at a redshift $z=0.01957$. The energy, timescale, host galaxy, and other properties of \src\ suggest that this source is most likely an off-axis long duration GRB. If so, the slow ($\sim t^{-1}$) decay of the radio afterglow confirms predictions for shock physics at late times \citep{ 2013ApJ...778..107S}. The energetics of the transient are also consistent with the ejecta from a binary neutron star merger, though the small host galaxy would be in tension with the hosts seen for short duration GRBs, which are typically $\sim 100$ times more massive. We also discuss the more speculative possibility that \src\ is a wind nebula produced in the aftermath of a magnetar-powered supernova. While the energy output of a newborn magnetar is poorly constrained by previous observations, such a scenario is also consistent with the luminosity, duration, and radio flux decay for \src. 

The detection of the radio afterglow of an orphan GRB would be the first of its kind \citep[see also][]{2013ApJ...769..130C, 2015ApJ...803L..24C} and improves the prospects for radio discovery of extragalactic transients. We note that \src\ was not identified as a transient or variable in comparisons of the FIRST and NVSS radio sky surveys \citep{2002ApJ...576..923L,2011ApJ...737...45O} because the two surveys happened to observe this slowly-evolving source within a few months of each other. This implies that published limits on radio transient event rates may be weaker than claimed for decades-long transients like \src. 

Furthermore, the brightness and proximity of \src\ implies that it has a relatively high volumetric rate, potentially in tension with predictions for orphan-GRB afterglows. We estimate that \src\ could have been detected if its peak brightness was as low as 4~mJy, equivalent to a 2.5$\times$\ larger distance. This implies that there may be an order of magnitude more such ``anti-transients'' to be found as FIRST sources that disappear in the VLASS and a comparable number of transients turning on in new observations. The chance of discovery will be improved with larger spectroscopic galaxy samples, better identification of non-nuclear radio sources, and integration of archival radio surveys with the search process. New searches for \src-like radio transients will improve the rate estimate and may reveal that it is inconsistent with that of orphan GRBs.

Independent of the progenitor model, \src\ is the oldest and best-localized luminous radio transient. That makes it a good place to search for remnants, since the ejecta should be transparent to free-free radiation. An FRB search may find bursts regardless of whether it is an GRB afterglow or magnetar wind nebula. Milliarcsecond-resolution imaging could distinguish GRB from magnetar models by measuring the size of the late-time radio emission. New measurements of the late-time radio flux of \src\ are needed to better measure its late-time radio decay and to properly search for scintillation. The presence of refractive scintillation at GHz frequencies is sensitive to spatial structure on a size scale of 10 to 100 $\mu$as and would help distinguish between the afterglow and magnetar wind nebula models.

\acknowledgments

We thank Rick Perley, Tom Osterloo, Jamie Farnes, Amy Kimball, Joe Callingham, Dustin Lang and Kevin Hurley for assistance with archival data, and Steve Croft, Geoff Bower, Xavier Prochaska, Dovi Poznanski and Hendrik Van Eerten for helpful comments.
C.J.L.\ acknowledges support from the National Science Foundation under grant 1611606. The Dunlap Institute is funded through an endowment established by the David Dunlap family and the University of Toronto. 
B.M.G.\ acknowledges the support of the Natural Sciences and Engineering Research Council of Canada (NSERC) through grant RGPIN-2015-05948, and of the Canada Research Chairs program.
B.D.M.\ acknowledges the support of NASA through the Astrophysics Theory Program through grant NNX16AB30G. 
E.O.O.\ is grateful for support by grants from the Israeli Ministry of Science, Minerva, BSF, BSF transformative program, Weizmann-UK, and the I-CORE Program of the Planning and Budgeting Committee and the Israel Science Foundation (grant no.\ 1829/12). 
L.S.\ acknowledges the support of NSF through grant AST-1716567.  The National Radio Astronomy Observatory is a facility of the National Science Foundation operated under cooperative agreement by Associated Universities, Inc.
The Pan-STARRS1 Surveys (PS1) and the PS1 public science archive have been made possirble through contributions by the Institute for Astronomy, the University of Hawaii, and others. Funding for the Sloan Digital Sky Survey IV has been provided by the Alfred P. Sloan Foundation, the U.S. Department of Energy Office of Science, and the Participating Institutions. SDSS acknowledges support and resources from the Center for High-Performance Computing at the University of Utah. The SDSS web site is \url{http://www.sdss.org}.
This research has made use of:
the SIMBAD database, operated at Centre de Donn\'{e}es astronomiques de
Strasbourg, France; the NASA/IPAC Extragalactic Database (NED) which is
operated by the Jet Propulsion Laboratory, California Institute of
Technology, under contract with NASA; NASA's Astrophysics Data System; and the VizieR catalog access tool, CDS, Strasbourg, France, as originally described by \citet{2000A&AS..143...23O}.

\vspace{5mm}
\facilities{VLA, EVLA, GMRT, WSRT, NCRTA, Beijing:MSRT, ATA, SDSS, Pan-STARRS}

\software{MIRIAD \citep{1995ASPC...77..433S}, astropy \citep{2018arXiv180102634T}, CASA \citep{2007ASPC..376..127M}, Aegean \citep{2012MNRAS.422.1812H,2018PASA...35...11H}, Afterglow Library \citep{2012ApJ...749...44V}.}

\bibliographystyle{aasjournal}

\end{document}